# Basics of averaging of the Maxwell equations

A. Chipouline, C. Simovski, S. Tretyakov

**Abstract**

Volume or statistical averaging of the microscopic Maxwell equations (MEs), i.e. transition from microscopic MEs to their macroscopic counterparts, is one of the main steps in electrodynamics of materials. In spite of the fundamental importance of the averaging procedure, it is quite rarely properly discussed in university courses and respective books; up to now there is no established consensus about how the averaging procedure has to be performed. In this paper we show that there are some basic principles for the averaging procedure (irrespective to what type of material is studied) which have to be satisfied. Any homogenization model has to be consistent with the basic principles. In case of absence of this correlation of a particular model with the basic principles the model could not be accepted as a credible one. Another goal of this paper is to establish the averaging procedure for metamaterials, which is rather close to the case of compound materials but should include magnetic response of the inclusions and their clusters. We start from the consideration of bulk materials, which means in the vast majority of cases that we consider propagation of an electromagnetic wave far from the interfaces, where the eigenwave in the medium has been already formed and stabilized. A basic structure for discussion about boundary conditions and layered metamaterials is a subject of separate publication and will be done elsewhere.

1. **Material equations representations**

We consider as a starting point a system of microscopic MEs in the following form:

$$\begin{cases} \operatorname{rot}\vec{e} = \dfrac{i\omega}{c}\vec{h} \\ \operatorname{div}\vec{h} = 0 \\ \operatorname{div}\vec{e} = 4\pi\rho \\ \operatorname{rot}\vec{h} = -\dfrac{i\omega}{c}\vec{e} + \dfrac{4\pi}{c}\vec{j} \\ \rho = \sum_i q_i \delta(\vec{r}-\vec{r}_i) \\ \vec{j} = \sum_i \vec{v}_i q_i \delta(\vec{r}-\vec{r}_i) \\ \dfrac{d\vec{p}_i}{dt} = q_i \vec{e} + \dfrac{q_i}{c}\left[\vec{v}_i * \vec{h}\right] \end{cases} \qquad (1)$$

Here $\vec{e}$ and $\vec{h}$ are the microscopic electric and magnetic fields, respectively, $\rho$ is the charge density, $\vec{q}_i$, $\vec{p}_i$, $\vec{r}_i$ and $\vec{v}_i$ are the charges, impulses, coordinates and velocities of charges, $\vec{j}$ is the microscopic current density, $\omega$ and $c$ are the frequency and the velocity of light in vacuum. It is assumed that system (1) is strictly valid without any approximations. Actually, system (1) can be elaborated in the framework of the minimum action approach [1]; nevertheless, one should remember that the minimum action principle does not give an unambiguous form of the MEs (1), but instead gives a set of different forms which satisfy the requirement of relativistic invariance. The "right" form can be chosen based on the evident requirement of correspondence of the results of the final system of equations to the observed physical effects. One should also mention that in the framework of the minimum action approach the final equations are written for "potentials + particles", not for "fields + particles"; the respective equations are:

$$\begin{cases} \dfrac{d\vec{p}_i}{dt} = -\dfrac{q}{c}\dfrac{\partial \vec{A}}{\partial t} - q_i \nabla \varphi + \dfrac{q_i}{c}\left[\vec{v}_i * \operatorname{rot} \vec{A}\right] \\ \dfrac{\partial F^{ik}}{\partial x^k} = -\dfrac{4\pi}{c} j^i, \quad F_{ik} = \dfrac{\partial A_k}{\partial x_i} - \dfrac{\partial A_i}{\partial x_k} \end{cases} \qquad (2)$$

Here $\vec{A}$ and $\varphi$ are the components of the 4-vector potential, and the relations between the microscopic fields and the potentials are given by:

$$\begin{cases} \vec{e} = -\dfrac{1}{c}\dfrac{\partial \vec{A}}{\partial t} - \nabla \varphi \\ \vec{h} = \operatorname{rot} \vec{A} \end{cases} \qquad (3)$$

Form (2) will not be used in the following discussions and is presented here just for methodological reasons.

We consider propagation of an electromagnetic plane wave interacting with the medium in case when the classical dynamics is supposed to be valid and the bulk material fills the whole space; the system (1) in this case can be formally averaged over a physically small volume (or through statistic averaging), which results in:

$$\begin{cases} \operatorname{rot}\langle\vec{e}\rangle = \dfrac{i\omega}{c}\langle\vec{h}\rangle \\ \operatorname{div}\langle\vec{h}\rangle = 0 \\ \operatorname{div}\langle\vec{e}\rangle = 4\pi\langle\rho\rangle \\ \operatorname{rot}\langle\vec{h}\rangle = -\dfrac{i\omega}{c}\langle\vec{e}\rangle + \dfrac{4\pi}{c}\langle\vec{j}\rangle \\ \rho = \sum_i q_i \delta(\vec{r}-\vec{r}_i) \\ \langle\vec{j}\rangle = \sum_i q_i \langle\vec{v}_i \delta(\vec{r}-\vec{r}_i)\rangle \\ \dfrac{d\vec{v}_i}{dt} = \dfrac{q_i}{m_i}\vec{e} + \dfrac{q_i}{m_i c}\left[\vec{v}_i * \vec{h}\right] \end{cases} \quad \begin{array}{c} \langle\vec{e}\rangle = \vec{E} \\ \\ \\ \longrightarrow \\ \\ \langle\vec{h}\rangle = \vec{B} \end{array} \quad \begin{cases} \operatorname{rot}\vec{E} = \dfrac{i\omega}{c}\vec{B} \\ \operatorname{div}\vec{B} = 0 \\ \operatorname{div}\vec{E} = 4\pi\langle\rho\rangle \\ \operatorname{rot}\vec{B} = -\dfrac{i\omega}{c}\vec{E} + \dfrac{4\pi}{c}\langle\vec{j}\rangle \\ \langle\rho\rangle = \sum_i q_i \delta(\vec{r}-\vec{r}_i) = \langle\rho\rangle(\vec{E},\vec{B}) \\ \langle\vec{j}\rangle = \sum_i q_i \langle\vec{v}_i \delta(\vec{r}-\vec{r}_i)\rangle = \langle\vec{j}\rangle(\vec{E},\vec{B}) \\ \dfrac{d\vec{v}_i}{dt} = \dfrac{q_i}{m_i}\vec{e} + \dfrac{q_i}{m_i c}\left[\vec{v}_i * \vec{h}\right] \end{cases} \qquad (4)$$

The averaging makes sense in case of a large number of atoms/molecules in the volume of averaging; from the other side the volume is supposed to be small in comparison with the wavelength of the electromagnetic wave, propagating in the medium.
The main problem after that is to find the averaged current and charge distribution as functions of the averaged electric and magnetic fields:

$$\langle \vec{j} \rangle = \langle \vec{j} \rangle (\vec{E}, \vec{B})$$
$$\langle \rho \rangle = \langle \rho \rangle (\vec{E}, \vec{B})$$
(5)

The last equation in (4) is not going to be averaged and describes the microscopic dynamics which is supposed to be substituted in $\langle \vec{j} \rangle$, $\langle \rho \rangle$ and averaged in order to get (5). Relations (5) in turn use information about microscopic dynamics as a function of microscopic fields which get averaged only after substitution into equations for $\langle \vec{j} \rangle$, $\langle \rho \rangle$. In fact, there is only one model (a multipole model [2]) where the averaging procedure for the $\langle \vec{j} \rangle$, $\langle \rho \rangle$ is performed rigorously, all other models do not even try to make this step and assume something phenomenological; the last equation in (4) is usually left out completely.

The system of equations (4), (5) is rather useless in practice until we make some progress and find analytical expressions for (5). Nevertheless, even without finding of an analytical form for (5), the averaged MEs can be analysed and important conclusions can be made.

It is worth noting that if we assume some analytical form for (5) (see, for example, [6] and references herein) then the averaging problem is basically fixed (or, better to say, bypassed), system (4) becomes self consistent and can be solved for the electric and magnetic fields $\vec{E}, \vec{B}$. All following below considerations (including introducing of $\vec{D}, \vec{H}$ in different representations, and permittivity and permeability) in this case are no more required. Thus, we assume in what follows that there are no explicit form of (5) and it is necessary to elaborate (5) further in order to find some reasonable analytical expressions for the averaged charge and current densities.

First, following [1] we consider a volume with charges and fields. The averaged charge in (5) can be represented through another function taking into account that the total charge of the considered volume is zero:

$$\int \langle \rho \rangle dV = 0$$
(6)

It means that the averaged density of charges can be presented as a divergence of another unknown function $\vec{P}_{full}$ (see more details in [1] and [8]):

$$\langle \rho \rangle = -\operatorname{div} \vec{P}_{full}$$
(7)

which is supposed to be zero outside the volume of integration in (6). In addition, this function is introduced with the accuracy of "rot" from any other arbitrary function $\vec{F}_1$:

$$\begin{cases} \langle \rho \rangle = -\operatorname{div} \vec{P}_{full} = -\operatorname{div}(\vec{P} + \operatorname{rot} \vec{F}_1) \\ \vec{P}_{full} = \vec{P} + \operatorname{rot} \vec{F}_1 \end{cases}$$
(8)

The averaged current is connected with the averaged charge density through the continuity relation [4], which remains valid for the macroscopic representation:

$$\begin{cases} i\omega\langle\rho\rangle = \operatorname{div}\langle\vec{j}\rangle \\ \langle\rho\rangle = -\operatorname{div}\vec{P}_{full} \end{cases} \quad (9)$$

This gives:

$$\operatorname{div}\left(\langle\vec{j}\rangle + i\omega\vec{P}_{full}\right) = 0 \quad (10)$$

The averaged current can be introduced with the accuracy of "rot" of one more arbitrary function $\vec{F}_2$:

$$\langle\vec{j}\rangle = -i\omega\vec{P}_{full} = -i\omega\vec{P}_{full} + \operatorname{rot}\vec{F}_2 \quad (11)$$

or, taking into account (8):

$$\langle\vec{j}\rangle = -i\omega\vec{P}_{full} + \operatorname{rot}\vec{F}_2 = -i\omega\vec{P} + \operatorname{rot}\left(-i\omega\vec{F}_1 + \vec{F}_2\right) \quad (12)$$

It turns out that the material equations (3) can be written through one new function $\vec{P}$ with the accuracy of two more arbitrary functions $\vec{F}_1$ and $\vec{F}_2$:

$$\begin{cases} \langle\rho\rangle = -\operatorname{div}\left(\vec{P} + \operatorname{rot}\vec{F}_1\right) \\ \langle\vec{j}\rangle = -i\omega\vec{P} + \operatorname{rot}\left(-i\omega\vec{F}_1 + \vec{F}_2\right) \end{cases} \quad (13)$$

It has to be mentioned that there are known concerns about possibility to present averaged density and current in form (13). In order to get (7) from (6) it is necessary to assume that the function $\vec{P}_{full}$ is zero outside the integration volume. In case of a medium with inclusions (which is the case for presented here discussion) the integration volume has to be taken inside the media and can cross the other volume of integration, where $\vec{P}_{full}$ is not zero. To avoid this contradiction, one can carefully choose the integration volume, which basically means, that the averaged characteristics depend on the choice of the integration volume, which should not be the case. Nevertheless, we can assume this situation for at least periodically spaced inclusions.

Another approach which is basically free from this drawback has been proposed in [5] and is called the scaling algorithm. The developed approach is based on a lemma proving that any field can be represented as a sum of three terms which are called "electric dipole", "magnetic dipole", and "electric quadrupole" moments (in the frequency domain):

$$J_i = -i\omega p_i + c e_{ijk}\frac{\partial}{\partial x_j}m_k - c\frac{\partial}{\partial x_k}\frac{\partial}{\partial t}q_{ik} = J_i^{(p)} + J_i^{(m)} + J_i^{(q)}$$

$$\begin{cases} m_i(x_j, J_k) = \dfrac{1}{2c}e_{ijk}x_j J_k \\ -i\omega q_{ij}(x_j, J_k) = \dfrac{1}{2c}(x_j J_i + x_i J_j) \\ -i\omega p_i(x_i, J_k) = -\left(x_j \dfrac{\partial}{\partial x_k} J_k\right) \end{cases} \tag{14}$$

This lemma leads to the possibility to represent, for example, the averaged current in the following form [5]:

$$\langle \vec{j} \rangle = -i\omega\left(\vec{P} - \operatorname{div}\hat{Q}\right) + c\operatorname{rot}\vec{M} \tag{15}$$

which basically repeats the second equation in (13). It is worth noting that equation (15) has been obtained without any additional limitations. The question about the physical meaning of the functions in (15) and (13) remains open.

Both functions $\vec{F}_1$ and $\vec{F}_2$ are arbitrary and independent. This means that it is up to us to impose any additional requirements on them. There are different but countable number of choices for the possible representations of (13). The most general case is when both $\vec{F}_1$ and $\vec{F}_2$ are non zero functions, namely:

$$\begin{cases} \vec{P}_{full} = \vec{P}_c = \vec{P} + \operatorname{rot}\vec{F}_1 \\ \vec{F}_2 = c * \vec{M}_C \end{cases} \tag{16}$$

which leads to the so called Casimir ("C") form of material equations:

$$\begin{cases} \langle \rho \rangle = -\operatorname{div}\vec{P}_C \\ \langle \vec{j} \rangle = -i\omega \vec{P}_C + c\operatorname{rot}\vec{M}_C \end{cases} \qquad \begin{cases} \vec{D} = \vec{E} + 4\pi\vec{P}_C \\ \vec{H} = \vec{B} - 4\pi\vec{M}_C \end{cases} \tag{17}$$

In this case MEs include four functions $\vec{E}, \vec{B}, \vec{D}, \vec{H}$:

$$\begin{cases} \operatorname{rot}\vec{E} = \dfrac{i\omega}{c}\vec{B} \\ \operatorname{div}\vec{B} = 0 \\ \operatorname{div}\vec{D} = 0 \\ \operatorname{rot}\vec{H} = -\dfrac{i\omega}{c}\vec{D} \end{cases} \tag{18}$$

Note, that the case $\vec{F}_1 = 0$ and $\vec{F}_2 = c*\vec{M}_C$ leads to the same form (16), where rot part of the full polarisability $\vec{P}_{full}$ is excluded (the physical mean of this part – presence of anapoles - will be considered below).

Alternative to (16), we can set:

$$\begin{cases} \vec{P}_{full} = \vec{P}_{LL} = \vec{P} + \text{rot}\,\vec{F}_1 \\ \vec{F}_2 = 0 \end{cases} \quad (19)$$

Which leads according to (13) to the so called Landau&Lifshitz ("L&L") form of material equations:

$$\begin{cases} \langle \rho \rangle = -\text{div}\,\vec{P}_{LL} \\ \langle \vec{j} \rangle = -i\omega \vec{P}_{LL} \end{cases} \qquad \begin{cases} \vec{D} = \vec{E} + 4\pi \vec{P}_{LL} \\ \vec{B} = \vec{B} \end{cases} \quad (20)$$

In this case MEs contain three functions $\vec{E},\ \vec{B},\ \vec{D}$:

$$\begin{cases} \text{rot}\,\vec{E} = \dfrac{i\omega}{c}\vec{B} \\ \text{div}\,\vec{B} = 0 \\ \text{div}\,\vec{D} = 0 \\ \text{rot}\,\vec{B} = -\dfrac{i\omega}{c}\vec{D} \end{cases} \quad (21)$$

Note that the form (20) does not assume that the averaged current $\langle \vec{j} \rangle$ does not contain any curl part – this part is included in $\langle \vec{j} \rangle$ through $\vec{F}_1$ (see (19)). The main difference between "C" and "L&L" representation is in the absence in the latter any stationary (not proportional to $\omega$) part of the curl part of $\langle \vec{j} \rangle$, described by $\vec{M}_C$. In case of the absence of stationary magnetization both representations have to be equivalent.

Finally, we assume that the full polarisability $\vec{P}_{full}$ contains only the curl part, namely:

$$\begin{cases} \vec{P}_A = \text{rot}\,\vec{F}_1 \\ \vec{F}_2 = c*\vec{M}_C \end{cases} \quad (22)$$

which leads, according to (13), to the case which we call here Anapole ("A") form of material equations:

$$\begin{cases} \langle \rho \rangle = 0 \\ \langle \vec{j} \rangle = -i\omega\,\text{rot}\,\vec{F}_1 + c\,\text{rot}\,\vec{M}_C = c\,\text{rot}\,\vec{M}_A \end{cases} \qquad \begin{cases} \vec{D} = \vec{E} + 4\pi\,\text{rot}\,\vec{F}_1 \\ \vec{H} = \vec{B} - 4\pi\vec{M}_A \end{cases}$$

In this case the system of MEs contains three functions $\vec{E}, \vec{B}, \vec{D}$ and reads:

$$\begin{cases} \operatorname{rot}\vec{E} = \dfrac{i\omega}{c}\vec{B} \\ \operatorname{div}\vec{B} = 0 \\ \operatorname{div}\vec{E} = 0 \\ \operatorname{rot}\vec{H} = -\dfrac{i\omega}{c}\vec{E} \end{cases} \qquad (24)$$

The physical object corresponding to such representation is an anapole [6], [8], which is now of great interest in connection with potential possibility of creation of such structures at nanoscales for optical wavelength region application [9].

Physical interpretation of the three mentioned above representations can be done based on the types of atoms/molecules (or meta-atoms/metamolecules) which the considered media consist of. In the most general case (dipole, quadrupole, dynamic and stationary magnetization, and anapoles) "C" form is preferable. In the case of absence of stationary magnetisation (but presence of all others) the "C" and "L&L" representations have to be equivalent. In case of absence of dipole, quadrupole, and magnetic dipole parts (presence of only anapoles and, maybe stationary magnetization) the "A" form is appropriate.

It is important to realize that there are no other choices for the material equations. Any homogenization model has to start from the statement in which representation it will be developed; arbitrary mixing between several representations is not acceptable, as it will be seen below.

Transformation between different forms of the representations is possible, taking into account mentioned above limitations for one or another representations. It is clear, that starting from "C" form (as the most general one) one can always get other three representations. From the other side, inverse transformations are not always possible – for example, "L&L" to "C" transformation assumes presence of stationary part of magnetization, which was not originally included in "L&L" form; the same is valid for "C" to "A" and "L&L" and "A" (and inverse) transformations due to the fact, that both "C" and "L&L" include anapole contributions, but "A" form does not include neither dipole/quadrupole nor magnetic dipole contributions.

Let us consider the relation between the forms of MEs, and start from the "C" form (17), (18). It is known that the "C" form is invariant with respect to the so called Serdyukov-Fedorov transformations:

$$\begin{cases} \vec{B}' = \vec{B} + \operatorname{rot}\vec{T_1} \\ \vec{E}' = \vec{E} - \dfrac{i\omega}{c}\vec{T_1} \\ \vec{D}' = \vec{D} + \operatorname{rot}\vec{T_2} \\ \vec{H}' = \vec{H} - \dfrac{i\omega}{c}\vec{T_2} \end{cases} \qquad (25)$$

Here $\vec{T}_1$ and $\vec{T}_2$ are arbitrary differentiable vector functions. The invariance means that for the new primed fields (22) the MEs keep their form (18).

The physical interpretation of the Serdyukov-Fedorov transformation is not trivial and will be considered elsewhere. Here it has to be pointed out that in these transformations the two pairs $\vec{E}, \vec{B}$ and $\vec{D}, \vec{H}$ are transformed independently, which obviously does not have too much physical sense. Nevertheless, here we (following [8]) apply these transformations rather formally and consider possible conclusions which can be made based on the application of the Serdyukov-Fedorov transformations.

For the material equations we can respectively write:

$$\begin{cases} \vec{D}' = \vec{E}' + 4\pi \vec{P}'_C \\ \vec{H}' = \vec{B}' - 4\pi \vec{M}'_C \\ \vec{D} = \vec{E} + 4\pi \vec{P}_C \\ \vec{H} = \vec{B} - 4\pi \vec{M}_C \\ \vec{P}'_C = \vec{P}_C + \dfrac{i\omega}{4\pi c} \vec{T}_1 - \dfrac{\text{rot}\,\vec{T}_2}{4\pi} \\ \vec{M}'_C = \vec{M}_C + \dfrac{i\omega}{4\pi c} \vec{T}_2 - \dfrac{\text{rot}\,\vec{T}_1}{4\pi} \end{cases} \qquad (26)$$

In order to get the ME for the new fields in "L&L" form we have to require that:

$$\begin{cases} \vec{H}' = \vec{B}' \\ \vec{M}'_C = \vec{M}_C + \dfrac{i\omega}{4\pi c} \vec{T}_2 - \dfrac{\text{rot}\,\vec{T}_1}{4\pi} = 0 \end{cases} \Rightarrow \vec{T}_2 = \dfrac{c}{i\omega}\left(\text{rot}\,\vec{T}_1 - \vec{M}_C\right) \qquad (27)$$

Substituting the last equation into (26), we finally have:

$$\begin{cases} \vec{D}' = \vec{E}' + 4\pi \vec{P}'_C \\ \vec{H}' = \vec{B}' \\ \vec{P}'_C = \vec{P}_C + \dfrac{i\omega}{4\pi c} \vec{T}_1 - \dfrac{c}{4\pi i\omega} \text{rot}\left(\text{rot}\,\vec{T}_1 - \vec{M}_C\right) \\ \vec{M}'_C = 0 \\ \vec{B}' = \vec{B} + \text{rot}\,\vec{T}_1 \\ \vec{E}' = \vec{E} - \dfrac{i\omega}{c} \vec{T}_1 \end{cases} \qquad (28)$$

which gives us the MEs in form of "L&L". If, in addition, we require that the electric and magnetic fields remain the same for both representations (which is reasonable, because both fields are assumed to be physically measurable), we obtain by setting $\vec{T_1}$ to zero:

$$\begin{cases} \vec{D'} = \vec{E} + 4\pi \vec{P'}_C \\ \vec{H'} = \vec{B'} = \vec{B} \\ \vec{P'}_{LL} = \vec{P_C} + \dfrac{c}{4\pi i\omega} \operatorname{rot} \vec{M_C} \\ \vec{M'}_{LL} = 0 \end{cases} \quad (29)$$

We see that starting from "C" representation, we can unambiguously reduce the MEs to the "L&L" form. It is important to emphasize, that in general both electric and magnetic fields are transformed and lose their initial physical means. The requirement of keeping the electric and magnetic fields the same in both representations is an additional one with respect to the Serdyukov-Fedorov transformation.

Let us consider the inverse transformation ("L&L" to "C" representation), namely we start from system (21) and write the Serdyukov-Fedorov transformation in this case:

$$\begin{cases} \vec{B'} = \vec{B} + \operatorname{rot} \vec{T_1} \\ \vec{E'} = \vec{E} - \dfrac{i\omega}{c} \vec{T_1} \\ \vec{D'} = \vec{D} + \operatorname{rot} \vec{T_2} \end{cases} \quad (30)$$

Substituting equations (30) into (21), we come to the conclusion that the MEs keep their form only if:

$$\vec{T_2} = -\dfrac{c}{i\omega} \operatorname{rot} \vec{T_1} \quad (31)$$

In this case we can formally assign:

$$\vec{B'} = \vec{H'} \quad (32)$$

and thus arrive to the "C" form. Nevertheless, in this case the electric field in new representation is not the same as in the old one – see (30). If we require again, as in previous case, that the electric and magnetic fields should not be changed at the transformations, we have to conclude that the "L&L" form is not invariant with respect to the Serdyukov-Fedorov transformation. Let us emphasize again, that starting from the "C" form it is possible to arrive to the "L&L" form, but starting from the "L&L" form it is impossible to reduce MEs to the "C" form using Serdyukov-Fedorov transformations and keeping electric and magnetic fields not transformed.

The inverse transformation is actually possible, if $\vec{P'}_{LL}$ clearly contains some curl part, in this case obvious separation of this part and grouping this part with $\vec{B}$ in MEs gives us immediately the MEs in "C" form. Note, that this transformation is not a part of the Serdyukov-Fedorov formalism and in general is not a trivial math problem. In general, one can use lemma (14) and present $\vec{P'}_{LL}$ as a sum of dipole, magnetic dipole, and quadrupole parts followed by mentioned above regrouping, which gives finally "C" form. Nevertheless, we have to remember that (14) is not constructive and does not give a clear recipe for such representation but rather proves its existence. Complete discussion of mutual "C" to "L&L" transformation is given in [8], where the equivalence between "C" and "L&L" form is given using consideration of boundary conditions, which is out of scope of this paper.

The direct and inverse transformations between "C" and "A", or between "L&L" and "A" forms can be considered in the same way as for "C" and "L&L" forms.

The presented above consideration of relations between the ""L&L" form and "C" form is very important. There is a commonly accepted integral form of $\vec{P}_{LL}$, which can be written according to the causality principle (which imposes limitations on the frequency dispersion form) and assuming that the physical processes at some point depend on the fields at other points (which gives rise to spatial dispersion):

$$\vec{P}_{LL}(\vec{k},\omega) = \chi_E(\vec{k},\omega)\vec{E}(\vec{k},\omega) + \chi_B(\vec{k},\omega)\frac{c}{i\omega}\left[\vec{k},\vec{E}(\vec{k},\omega)\right] \tag{33}$$

where the first term is responsible for interactions with the electric field, and the second one – with the magnetic field. The last term is often omitted assuming that the constant takes into consideration both contributions:

$$\vec{P}_{LL}(\vec{k},\omega) = \chi(\vec{k},\omega)\vec{E}(\vec{k},\omega) \tag{34}$$

When we use this representation, it has to be clearly realized that:

1. We are working with the "L&L" representation where there is no magnetization (the magnetic response is included through the spatial dispersion of electric polarization).
2. The form (34) is NOT relativistic invariant, because of it DOES NOT take into account retardation effect in spatial dispersion; practically it means that the (34) makes sense for small $\vec{k}$ only.
3. The form (33) and (34) assume translation invariance of the media, which means that the form is acceptable for homogeneous material far from the boundaries. Thus the representation (33) and (34) can NOT be used as a phenomenological form to describe boundaries, for instance reflection properties. The form (33) and (34) can only be used as a basis for dispersion relation inside the materials.
4. Starting from (34), it is impossible to introduce any permeability, because of in "L&L" representation there is no magnetization. All magnetic effects are included in the permittivity. In order to introduce a permeability it is necessary to transit to "C" form taking into account the mentioned above restrictions.
5. Basically, there is no reason to introduce permittivity or/and permeability provided there are some analytic forms for the averaged charge and current densities $\langle\vec{j}\rangle$, $\langle\rho\rangle$, because of known $\langle\vec{j}\rangle$, $\langle\rho\rangle$ as functions of the averaged fields (5) closes the system of MEs. The permittivity and permeability can nevertheless be introduced (through material equations) but they cannot in general be used in order to describe boundary effects (transmission and reflection), because in the vicinity of boundary the permittivity and permeability differ from their bulk counterparts.

All the considerations above did not answer the question "How to get the unknown functions for the polarizability ("L&L" form) or polarizability and magnetization ("C" form)?" starting from the microscopic picture.

The main problem is to develop a model, which would give us a recipe to find expressions for $\vec{P}$ and $\vec{M}$ in (17) or (20) as a functions of the averaged fields – it has to be also pointed out, that the expressions have to be presented as functions of the averaged (macroscopic), and not the microscopic fields; only in this case we can formulate the MEs as a self consistent system. Nevertheless, it is important to realize, that whatever model is developed, it can be presented in "C", or in "L&L", or in "A" form with the respective consequences, described above, and the difference between the bulk and boundary situations has to be clearly distinguished.

## 2. Models for polarization P and magnetization M

Introduction of the unknown functions $\vec{P}$ and $\vec{M}$ are made following mainly two different ways, namely:

1. By introducing multipole moments (which will be call below the <u>Multipole model</u>), which leads finally to the "C" form (as it has been shown above, in this case the "L&L" form can be obtained using the Serdyukov-Fedorov transformations).
2. By introduction of the phenomenological integral (31) for polarization (which will be called below the <u>Phenomenological model</u>). The "L&L" form of representation is evidently preferable in this case, because there is only one unknown function $\vec{P}_{LL}$, which is supposed to contain all the information about intrinsic material properties (material equations are expressed through this function). As a result, we are working in the frame of the "L&L" approach, and transformation to the "C" representation (for example, to introduce permeability) is not straightforward.

### 2.1 Multipole model

The Multipole model has been put forward in [2], and then developed in a similar form in [3]. The model results in expressions for $\vec{P}_C$ and $\vec{M}_C$ presented through the averaged dynamics of the charges:

$$\begin{cases} \vec{P}_C(\vec{R},\omega) = \left\langle \sum_s^{all\,charges} e_s r_s \right\rangle - \nabla \bullet Q_e \\ Q_{ij}(\vec{R},\omega) = \frac{1}{2} \left\langle \sum_s^{all\,charges} q_s r_{i,s} r_{j,s} \right\rangle \\ \vec{M}_C(\vec{R},\omega) = \frac{1}{2c} \left\langle \sum_s^{all\,charges} \left[ \vec{r}_s, q_s \frac{d\vec{r}_s}{dt} \right] \right\rangle \end{cases} \quad (35)$$

Here it is important to realize that the formulas for the macroscopic polarization and magnetization are expressed in averaged dynamics, which are a priori functions of microscopic (not macroscopic) fields. Even if we are able to write analytical forms for the dynamics, we will have to express the microscopic fields through the macroscopic ones, which returns us to the main problem of all this consideration.

**2.2 Phenomenological model**

The Phenomenological model (34) is widely used in different branches of physics like plasma physics or physics of crystals. In the vast majority of the considered problems, expression (34) is expanded into the Taylor series up to the second order, namely:

$$\vec{P}_{LL}(\vec{k},\omega) = \chi(\vec{k},\omega)\vec{E}(\vec{k},\omega) \approx \left[\chi(\vec{k}_0,\omega) + \frac{\partial \chi(\vec{k}_0,\omega)}{\partial \vec{k}}\bigg|_{\vec{k}=\vec{k}_0}(\vec{k}-\vec{k}_0) + \frac{1}{2}\frac{\partial^2 \chi(\vec{k}_0,\omega)}{\partial \vec{k}^2}\bigg|_{\vec{k}=\vec{k}_0}(\vec{k}-\vec{k}_0)^2\right]\vec{E}(\vec{k},\omega)$$

(36)

The coefficients $\chi(\vec{k}_0,\omega)$, $\frac{\partial \chi(\vec{k}_0,\omega)}{\partial \vec{k}}\bigg|_{\vec{k}=\vec{k}_0}$, $\frac{\partial^2 \chi(\vec{k}_0,\omega)}{\partial \vec{k}^2}\bigg|_{\vec{k}=\vec{k}_0}$ are supposed to be found from experiments or rigorous microscopic calculations.

Concerning (36) several things have to be pointed out, namely:

1. Expansion can be performed around any $\vec{k}_0$, not necessarily $\vec{k}_0 = 0$, provided expansion over angles and wavelengths is properly done; in other words, the expansion formally can be written for small spatial dispersion and for large spatial dispersion. Math in this case DOES NOT impose any limitations.
2. Form (31) assumes no retardation. In order to keep the physical meaning for this form we have to require that (31) has to be applied only in cases where retardation does not play any role; mathematically it means that we consider the volumes smaller than the wavelengths of interest, which obviously corresponds to the case $\vec{k}_0 = 0$ in expansion (33).
3. It is worth noting, that in spite of the requirement $\vec{k}_0 = 0$ expansion (33) DOES NOT assume independence of material properties on the propagation direction!!! In general, this dependence is contained in the expansion coefficients; even though they are taken at $\vec{k}_0 = 0$!
4. It should be again emphasized that form (33) as well as (30) and (31) can be used only to describe properties inside the material, where the translational invariance is satisfied. Any questions concerning boundaries can NOT be considered based on this form and require additional models of the boundary properties.

After having all this said, the final form of the $\vec{P}_{LL}$ for "L&L" representation can be written as:

$$\vec{P}_{LL}(\vec{k},\omega) = \chi(\vec{k},\omega)\vec{E}(\vec{k},\omega) \approx \left[\chi(\omega) + \frac{\partial \chi(\vec{k}_0,\omega)}{\partial \vec{k}}\bigg|_{\vec{k}=0}\vec{k} + \frac{1}{2}\frac{\partial^2 \chi(\vec{k}_0,\omega)}{\partial \vec{k}^2}\bigg|_{\vec{k}=0}(\vec{k})^2\right]\vec{E}(\vec{k},\omega) \quad (37)$$

**2.3 Dispersion relation and introduction of permittivity and permeability for bulk and boundaries**

It appears commonly in the literature to start discussion of the material properties from some expressions for dielectric (permittivity) and magnetic (permeability) constants. This approach has been used in the first papers about negative refractive index in metamaterials, this approach is used in plasma physics, etc. However, methodologically it is not quite well founded, and it would be reasonable to clarify the situation with introducing of permittivity and permeability.

First, let us consider the problem of propagation of electromagnetic waves in a bulk infinite medium (without consideration of boundary conditions). In this case the solution of the problem can be expressed in form of plane waves, and the main result is (in case of the linear problem) the dispersion relation $\vec{k}(\omega)$. The dispersion relation is in turn given by the system of field equations (for the "C" and "L&L" representation respectively):

$$\begin{cases} \text{rot}\,\vec{E} = \dfrac{i\omega}{c}\vec{B} \\ \text{div}\,\vec{B} = 0 \\ \text{div}\,\vec{D} = 0 \\ \text{rot}\,\vec{H} = -\dfrac{i\omega}{c}\vec{D} \\ \vec{D} = \vec{E} + 4\pi\vec{P}_C \\ \vec{H} = \vec{B} - 4\pi\vec{M}_C \end{cases} \quad (38a) \qquad \begin{cases} \text{rot}\,\vec{E} = \dfrac{i\omega}{c}\vec{B} \\ \text{div}\,\vec{B} = 0 \\ \text{div}\,\vec{D} = 0 \\ \text{rot}\,\vec{B} = -\dfrac{i\omega}{c}\vec{D} \\ \vec{D} = \vec{E} + 4\pi\vec{P}_{LL} \end{cases} \quad (38b)$$

As we see, there are absolutely NO reasons for introduction some new constants, especially if both $\vec{P}$ and $\vec{M}$ in the "C" representation depend on both electric and magnetic fields; in case of the "L&L" representation it is even less justified.

The permittivity and permeability are introduced in this case in order to avoid microscopic considerations for the $\vec{P}$ and $\vec{M}$. For the case of the "L&L" representation, for example, it is often assumed that (see, for example, [8]):

$$\varepsilon_{ij}(\vec{k},\omega) = \varepsilon^{tr}(\omega)\left(\delta_{ij} - \frac{k_i k_j}{k^2}\right) + \varepsilon^{l}(\omega)\frac{k_i k_j}{k^2} \tag{39}$$

which results in dispersion relation:

$$k^2(\omega) = \frac{\omega^2}{c^2}\varepsilon(\vec{k},\omega) \tag{40}$$

Equation (40) can be solved with respect to $\vec{k}(\omega)$ which finally gives us all branches of $\vec{k}(\omega)$. After substituting $\vec{k}(\omega)$ into (39) one can write down expression

$$\varepsilon_{ij}(\vec{k}(\omega),\omega) = \varepsilon_{ij}(\omega) \tag{41}$$

without any apparent spatial dispersion. It indicates that in this formalism the spatial dispersion is no more than a math tool to describe the properties of the $\vec{P}$ and $\vec{M}$ which is useful in some cases, and which physically stems from nonlocal medium response. Finally, for a particular dispersion curve (in case of multiple dispersion curves) the dielectric constant is written without any spatial dispersion.

It is worth noting that form (36) does not give us any recipe for finding unknown $\varepsilon(\omega)$ (or $\varepsilon(\omega)$ and $\mu(\omega)$ in case of the "C" representation). Introduction of $\varepsilon(\omega)$ and $\mu(\omega)$ is no more than a phenomenological introduction of a relation of $\vec{P}$ and $\vec{M}$ from the averaged field (5), which is basically the key question in homogenization. Bypassing this problem by a phenomenological introduction of a relation for $\vec{P}$ and $\vec{M}$ (5) means that basically the main problem of the homogenization is bypassed as well.

Form (39) is no more than a phenomenological approach for writing expressions (17) and (20) without creation of some model which would give us $\vec{P}$ and $\vec{M}$; in contrast, system (32) creates a transition between the microscopic and macroscopic pictures. It is also important to mention that (35) is applicable for both bulk materials and boundary problems (provided that the averaging is done correctly) in contrast with the phenomenological approach (33), (34), (37), which is restricted naturally by consideration of plane waves in bulk infinite materials.

It should be realized also that the permittivity and permeability, which can be introduced for bulk materials can NOT be apiori used for modelling boundary phenomena, for example for reflection and transmission calculation based on the Fresnel equations. The permittivity and permeability in general will be different in the bulk and close to boundaries, which can be understood from (32) – the averaging near a boundary is definitively NOT the same as inside the bulk material.

### 3. Conclusions

As conclusions:
1. Macroscopic, averaged MEs (more rigorously, material equations) can be presented in three different forms, namely "C", "L&L", and "A" forms.
2. Starting from "C" form, one can unambiguously define the "L&L" and "A" forms.
3. Starting from "L&L" or "A" form, in general it is impossible to define "C" form.
4. Starting from "L&L" form, it is possible to derive "A" form, starting from "A" form it is generally impossible to derive neither "L&L" nor "C" form.
5. In any consideration, one has to define in which representation - "C", "L&L", or "A" – a particular discussion is going on.
6. The result of the averaging for bulk material is expressed in form of dispersion relation, not in form of some functional forms for permittivity and permeability; moreover, there are no reasons to introduce them. Permittivity and permeability can be introduced phenomenological or in the microscopic way in order to write some expressions for $\vec{P}$ and $\vec{M}$ in (17) and (20) which are used to elaborate the dispersion relation.